# Synthesis, Crystal Structure, and Thermoelectric Properties of Layered Antimony Selenides REOSbSe$_2$ (RE = La, Ce)

Yosuke Goto,[1]* Akira Miura,[2] Ryosuke Sakagami,[3] Yoichi Kamihara,[3] Chikako Moriyoshi,[4] Yoshihiro Kuroiwa,[4] and Yoshikazu Mizuguchi[1]

[1]*Department of Physics, Tokyo Metropolitan University, 1-1 Minami-osawa, Hachioji, Tokyo 192-0397, Japan*
[2] *Faculty of Engineering, Hokkaido University, Kita 13, Nishi 8, Sapporo 060-8628, Japan*
[3] *Department of Applied Physics and Physico-Informatics, Faculty of Science and Technology, Keio University, 3-14-1 Hiyoshi, Yokohama 223-8522, Japan*
[4] *Department of Physical Science, Hiroshima University, 1-3-1 Kagamiyama, Higashihiroshima, Hiroshima 739-8526, Japan*

Inspired by the recent first-principles calculations showing the high thermoelectric performance of layered pnictogen chalcogenides, we experimentally characterise the crystal structure and high-temperature transport properties of the layered antimony selenides REOSbSe$_2$ (RE = La, Ce). The crystal structure of REOSbSe$_2$ was the tetragonal *P*4/*nmm* space group, consisting of alternate stacks of SbSe$_2$ and REO layers. These two compounds were n-type semiconductors. The optical band gaps of LaOSbSe$_2$ and CeOSbSe$_2$ were evaluated to be 1.0 and 0.6 eV, respectively. The room-temperature thermal conductivity was 1.5 Wm$^{-1}$K$^{-1}$ for RE = La and 0.8 Wm$^{-1}$K$^{-1}$ for RE = Ce. These relatively low thermal conductivities were comparable to those of isostructural layered bismuth chalcogenides. We substituted O$^{2-}$ with F$^-$ ions to introduce electrons as charged carriers to optimize the thermoelectric performance, but increasing the electrical conductivity was still challenging.

## 1. Introduction

Searching for highly efficient thermoelectric materials that convert temperature difference into electricity is of crucial importance for energy harvesting from waste heat.[1–5)] The performance of thermoelectric materials is governed by the dimensionless figure of merit, *ZT* = $S^2T\rho^{-1}\kappa^{-1}$, where *T* is the absolute temperature, *S* is the Seebeck coefficient, $\rho$ is the electrical resistivity, and $\kappa$ is the thermal conductivity. These transport properties are highly dependent on the electronic band structure as well as the carrier concentration, and the development of materials that exhibit high *ZT* in a broad temperature range has been desired. Recent advances in the first-principles approach have enabled reliable predictions of the electronic structures and transport properties under some assumptions, such as a constant relaxation time.[6–12] Indeed, several candidate materials have been theoretically predicted as efficient thermoelectric materials, and the performances of some of these predicted compounds were experimentally demonstrated.

Recently, Ochi et al. presented first-principles calculations for the electronic structure and carrier transport of layered pnictogen chalcogenides REOPnCh$_2$ (RE = rare-earth ion, Pn = pnictogen, Ch = chalcogen).[13] This family of compounds has been extensively studied as a superconductor when Pn = Bi.[14–16] The crystal structure of these compounds is characterized by alternate stacks of PnCh$_2$ conducting layers and REO spacer layers, as schematically shown in Fig. 1. Regarding the thermoelectric performance, $ZT$ = 0.36 at 650 K was reported for LaOBiSSe.[17] Although $ZT$ for this family of compounds is still moderate, their characteristics, namely, (1) intrinsically low lattice thermal conductivity of 1–2 Wm$^{-1}$K$^{-1}$,[18–21] and (2) a large degree of freedom regarding elemental substitution and carrier doping, indicate their promise as an efficient thermoelectric system. In particular, a decrease in lattice thermal conductivity due to the rattling motion of Bi was recently reported, even though these compounds have no oversized cage.[22] Therefore, improving the electrical power factor ($S^2\rho^{-1}$) is essential to develop efficient thermoelectric materials based on the layered pnictogen chalcogenides. Ochi et al. presented the guiding principles for improving the power factor of the LaOBiS$_2$-type compounds: (1) small spin-orbit coupling, (2) small Pn–Pn and large Pn–Ch hopping amplitudes, and (3) a small onsite energy difference between the Pn–$p_{xy}$ and Ch–$p_{xy}$ orbitals.[13] Consequently, replacements of Bi and S ions with lighter and heavier elements, respectively, should result in increased thermoelectric performance. It was predicted that REOAsSe$_2$ will show $ZT$ exceeding 2 under some plausible assumptions. Unfortunately, our preliminary experiments on the synthesis of REOAsSe$_2$ have not yet been successful because the RE$_4$O$_4$Se$_3$ phase is obtained as a stable phase.

In the present study, we present the synthesis, crystal structure, and thermoelectric properties of REOSbSe$_2$ (RE = La, Ce). The crystal structure of LaOSbSe$_2$ has been reported to be the tetragonal $P4/nmm$ space group,[23] and a theoretical study showed that electrically tunable multiple Dirac cones exist in LaOSbSe$_2$ thin films.[24] However, detailed physical properties of LaOSbSe$_2$ are yet to be reported. CeOSbSe$_2$, which was a newly obtained phase in this work, was found to be isostructural with LaOSbSe$_2$. We found that both LaOSbSe$_2$ and CeOSbSe$_2$ are n-type semiconductors with direct-type optical band gaps of 1.0 eV for RE = La and 0.6 eV for RE = Ce. Relatively low thermal conductivity of ~1 Wm$^{-1}$K$^{-1}$ was observed, similar to that of the related compounds LaOBiS$_{2-x}$Se$_x$. We also described the effects of carrier doping by the partial substitution of O$^{2-}$ with F$^-$ ions.

## 2. Experimental Details

Polycrystalline samples were prepared via the two-step solid-state reactions established for the synthesis of iron-based oxypnictides by Kamihara et al.[25,26] LaOSbSe$_2$ was synthesized using dehydrated La$_2$O$_3$, a mixture of compounds mainly composed of LaSe and LaSe$_2$ (LaSe–LaSe$_2$ powder), Sb (99.9%), and Se (99.999%) as starting materials. The dehydrated La$_2$O$_3$ was prepared by heating commercial La$_2$O$_3$ powder (99.9%) at 600 °C for 10 h in air. To obtain the LaSe–LaSe$_2$ powder, La (99.9%) and Se were mixed in a ratio of 2 : 3 and heated at 500 °C for 10 h in an evacuated silica tube. Because La powder is reactive in air and a moist atmosphere, this process was carried out in an Ar-filled glovebox. Then, a mixture of these starting materials was pressed into a pellet and heated at 700 °C for 20 h in an evacuated silica tube. The obtained product was ground, mixed, pelletized, and heated again under the same heating conditions. To obtain dense samples, the powder samples were hot-pressed using a graphite die at 700 °C at 50 MPa for 30 min (S. S. Alloy, PLASMAN CSP-KIT-02121). The hot-pressed samples had a relative density greater than 97%. To obtain a F-doped sample with a nominal composition of LaO$_{0.9}$F$_{0.1}$SbSe$_2$, LaF$_3$ (99.9%) powder was added to the starting materials. CeOSbSe$_2$ and CeO$_{0.9}$F$_{0.1}$SbSe$_2$ were also prepared via the above synthesis process, employing CeO$_2$ (99.99%), Ce (99.9%), Sb, Se, and CeF$_3$ (99.9%) as starting materials.

The phase purity and crystal structure of the samples were examined using synchrotron powder X-ray diffraction (SPXRD) performed at the BL02B2 beamline of SPring-8 (proposal number 2017B1211). The diffraction data were collected using a high-resolution one-dimensional semiconductor detector (MYTHEN).[27] The wavelength of the radiation beam was determined to be 0.495586(1) Å using a CeO$_2$ standard. The crystal structure parameters were refined using the Rietveld method using RIETAN-FP software.[28] The crystal structure was visualized using VESTA software.[29]

The sample morphology and chemical composition were examined using a scanning electron microscope (SEM; Hitachi, TM3030) equipped with an energy-dispersive X-ray spectrometer (EDX; Oxford, SwiftED3000). Diffuse reflectance ($R$) spectra were collected using a spectrometer equipped with an integrating sphere (Hitachi, U-4100) and were converted into absorption ($\alpha$) spectra using the Kubelka−Munk relationship, $\alpha/s =$

$(1-R)^2/2R$, where $s$ is the scattering factor.[30]

The electrical resistivity $\rho$ and Seebeck coefficient $S$ were simultaneously measured using the conventional four-probe geometry (Advance Riko, ZEM-3) in a He atmosphere. Typically, samples were $2 \times 3 \times 9$ mm$^3$ in size and rectangular parallelepiped in shape. In this study, the measurement results in both heating and cooling process were obtained. The thermal conductivity was calculated using $\kappa = DC_p d_s$, where $D$, $C_p$, and $d_s$ are the thermal diffusivity, specific heat, and sample density, respectively. The thermal diffusivity was measured by a laser flash method (Advance Riko, TC1200-RH). The samples used for the measurements were disks 10 mm in diameter and 2 mm in thickness. The specific heat was measured by a comparison method using differential scanning calorimetry (DSC; Advance Riko, DSC-R) under an Ar atmosphere.

## 3. Results and Discussion

### 3.1 Structural characterization

Figure 2 shows the SPXRD pattern and Rietveld fitting results for LaOSbSe$_2$ and CeOSbSe$_2$. Almost all the diffraction peaks can be assigned to those of the tetragonal *P*4/*nmm* space group, except for several tiny peaks attributable to Sb (1.3 wt%) and La$_2$O$_2$Se (1.5 wt%) in LaOSbSe$_2$, and CeO$_2$ (4.1 wt%) in CeOSbSe$_2$. The chemical composition evaluated using EDX is summarized in Table S1 in the Supporting Information.[31] The results show that the off-stoichiometry of the samples was less than the uncertainty of EDX, which is typically about 5%. We also analyzed the crystal structure using the structure model containing defects. However, it did not improve the refinement, confirming that the samples were nearly stoichiometric. The results of the Rietveld analysis including the refined structural parameters are listed in Table I and II, and the selected bonding distances and angles of LaOSbSe$_2$ are depicted in Fig. 1(b). The lattice parameters were $a = 0.414340(3)$ nm and $c = 1.434480(14)$ nm for LaOSbSe$_2$, and $a = 0.408624(3)$ nm and $c = 1.417697(15)$ nm for CeOSbSe$_2$. The decrease in the lattice parameters of the Ce analogue is most likely due to the smaller ionic radius of Ce ions (Shannon's six-coordinate ionic radius, $r_{Ce}^{3+} = 101$ pm and $r_{Ce}^{4+} = 87$ pm) than that of La ions ($r_{La}^{3+} = 103.2$ pm).[32] A bond valence sum calculation for the RE site shows that the valence state of La in LaOSbSe$_2$ is +3.06, while that of Ce in CeOSbSe$_2$ is +3.29.[33,34] The results indicate that Ce ions in CeOSbSe$_2$ exhibit a mixed valence of 3+ and 4+. Indeed, in the case of the layered bismuth chalcogenides, it has been reported that Ce ions act as an electron-carrier dopant because of the mixed valence.[35–39] Generally, Curie–Weiss analysis of the magnetization as a function of temperature is a typical and effective way to estimate the valence state of Ce ions. However, it is not an easy task to precisely determine the Ce valence using this analysis because our sample contains $CeO_2$ as an impurity. It has been reported that $CeO_2$ clusters may significantly contribute to the measured magnetic behavior, and the origin of the anomalous magnetization of $CeO_2$ is still under debate.[40]

The lattice volume of the F-doped samples, nominally $LaO_{0.9}F_{0.1}SbSe_2$ and $CeO_{0.9}F_{0.1}SbSe_2$, was smaller than that of the undoped samples, where the SPXRD patterns are shown in Fig. S1 in the Supporting Information. For the RE = La sample, the diffraction peaks were broadened as a result of $F^-$ substitution, suggesting a decrease in crystallinity. In contrast, the diffraction peaks of $CeO_{0.9}F_{0.1}SbSe_2$ are still relatively sharp, although our preliminary experiments showed that the solubility limit of $F^-$ ions is around 10% for both $LaOSbSe_2$ and $CeOSbSe_2$ (data not presented).

For the sites of Sb and in-plane Se (Se1, see Fig. 1), we refined the anisotropic atomic displacement parameters (ADPs) because these parameters may be essential to the electrical transport in the phases. In the case of the related system of layered bismuth chalcogenides, the in-plane ADPs were reported to be an indicator of in-plane disorder and are correlated with the evolution of superconductivity.[41,42] As shown in Fig. 1(b), the Se1 sites of these compounds exhibited relatively large displacements in the *ab*-plane. These large ADPs are probably related to the difficulty of synthesizing conductive samples, as demonstrated later.

We examined the sample morphology on a polished surface and pulverized powder using a SEM. As shown in Fig. 3, the SEM image of the polished surface confirms the high density of the samples, and the diameter of the secondary phase and voids was ~1 μm. The primary particle (grain) size of $LaOSbSe_2$ was also ~1 μm. We note that the particle size of $LaO_{0.9}F_{0.1}SbSe_2$ was comparable to that of undoped $LaOSbSe_2$, although its X-ray diffraction pattern showed broadened peaks (see Fig. S2).

### 3.2 Optical properties

Figure 4(a) shows ($\alpha$/s) plotted as a function of photon energy. The direct-type absorption edge was evaluated to be 1.0 eV for $LaOSbSe_2$ and 0.6 eV for $CeOSbSe_2$ from the onset of the $(\alpha h\nu/s)^2$ plot, as shown in Fig. 4(b). These results are interesting because it is believed that the conduction band minimum (CBM) and valence band maximum (VBM) of these compounds are mainly composed of hybridization between Sb and Se orbitals, and the RE element has a minor effect on the electronic structure close to the CBM and VBM.[13,43)] The reduced band gap in CeOSbSe$_2$ suggests that the band width of the conduction band and valence band was increased owing to the increased overlap between the Sb-p and Se1-p orbitals by chemical pressure effects. It might be possible that a Ce 4f orbital creates an energy level between the CBM and VBM to reduce the band gap. Photoemission spectroscopy will be helpful to precisely determine the Ce 4f energy level in CeOSbSe$_2$.[37)]

**3.3 Transport properties**

Figure 5(a) shows the temperature dependence of the electrical resistivity. For LaOSbSe$_2$, the room-temperature electrical resistivity is as high as $10^3$ Ωm, and it decreases with increasing temperature. As shown in Fig. 5(b), the $\rho$–$T$ data above 300 K is well described using an Arrhenius plot with a thermal activation energy ($E_a$) of 0.43 eV. This is close to half of the optical band gap, 1.0 eV, suggesting LaOSbSe$_2$ can be regarded as an intrinsic semiconductor. Electrical resistivity was distinctly lowered as a result of F substitution: $\rho \sim 1$ Ωm at 300 K and $E_a = 0.22$ eV for LaO$_{0.9}$F$_{0.1}$SbSe$_2$ (the Arrhenius plot is shown in Fig. S3). However, these samples are still highly resistive for a thermoelectric material. For CeOSbSe$_2$, the room-temperature electrical resistivity is 10 Ωm, which is two orders of magnitude lower than that of LaOSbSe$_2$. This is most likely due to the increased carrier density via the mixed valence of Ce ions and/or reduced band gap, as described above. At the same time, the electrical resistivity of CeO$_{0.9}$F$_{0.1}$SbSe$_2$ is almost equal to that of undoped CeOSbSe$_2$. This is probably because the mixed valence of Ce ions was suppressed in the F-doped sample. In related systems containing mixed-valent RE ions, including CeO$_{1-x}$F$_x$BiS$_2$, Eu$_{0.5}$La$_{0.5}$FBiS$_{2-x}$Se$_2$, and La$_{1-x}$Ce$_x$OBiSSe, the mixed valence of the RE ions (Eu or Ce) is sensitive to the local crystal structure,[36,37,39,44)] which is qualitatively described in terms of the in-plane chemical pressure effect. Because F substitution with O ions decreases the lattice parameters and thus increases the in-plane chemical pressure, the suppression of the mixed valence of Ce ions for CeO$_{0.9}$F$_{0.1}$SbSe$_2$ seems to be consistent with the observed results in the bismuth chalcogenide systems. Figure 6 shows the temperature dependence of the Seebeck coefficient. The negative value of the observed Seebeck coefficient indicates that the majority carrier is electrons in all samples. Generally speaking, the measured results of the Seebeck coefficient are consistent with that of the electrical resistivity; the sample showing lower resistivity exhibits a lower absolute value of the Seebeck coefficient as a result of increased carrier density. It is noteworthy that the absolute value of the Seebeck coefficient decreases with increasing temperature, suggesting thermally excited holes also contribute to the carrier transport, as well as electrons. For $LaO_{0.9}F_{0.1}SbSe_2$, the temperature dependence of the Seebeck coefficient was slightly suppressed. This is due to the increase in the Fermi level caused by F substitution, as demonstrated by the decreased activation energy.

Figure 7 shows the temperature dependence of the thermal conductivity. For $LaOSbSe_2$, the room-temperature thermal conductivity was 1.5 $Wm^{-1}K^{-1}$, which is comparable to the results of related compounds, 1–2 $W m^{-1}K^{-1}$.[18–21] Interestingly, the thermal conductivity of the samples was reduced to <1 $W m^{-1}K^{-1}$ for $CeOSbSe_2$ and the F-doped samples. Because the electrical component of the thermal conductivity estimated using the Wiedemann–Franz relationship was less than 1% of the measured thermal conductivity, the decrease in thermal conductivity should be attributed to phonon transport properties. One possible explanation of this reduced thermal conductivity is the strained and distorted crystal structure of the layered mixed-anion compounds.[2,5,22] Although detailed insights into the phonon properties of these compounds are beyond the scope of the present study, we deduce that this reduced thermal conductivity is correlated with the unusual lattice vibration. In the case of the layered bismuth chalcogenides $LaOBiS_{2-x}Se_x$, the decrease in lattice thermal conductivity is correlated with the Bi vibration energy, namely, the rattling motion of Bi ions with no oversized cage.[22] Such a mechanism might also occur in the present Sb-based systems. A detailed study combined with inelastic neutron scattering measurements and first-principles calculations will shed light on the phonon properties in the present systems.

We briefly discuss the related systems. For example, $SbS_2$-based compounds are also potentially high-performance thermoelectric materials according to a theoretical calculation by Ochi et al.[13] However, our preliminary experiments showed that $NdOSbS_2$ is an insulator, as in the case of $Ce(O,F)SbS_2$.[45] Although $LaOSbSe_2$ and $CeOSbSe_2$ examined in this study are still highly resistive, they show semiconducting transport, most likely due to the reduced band gap with respect to the $SbS_2$-based system. Therefore, we believe that the $SbSe_2$-based system is suitable to experimentally assess the theoretical prediction by Ochi et al. Our preliminary experiments also showed that the solubility limit of fluorine strongly depends on the RE ion, suggesting the synthesis of $SbSe_2$-based materials with sufficient conductivity is possible. Detailed studies of the crystal structure and transport properties of other $SbSe_2$-based compounds are currently under way.

## 4. Summary

We synthesized the layered antimony selenides $LaOSbSe_2$ and $CeOSbSe_2$ to examine their crystal structure and transport properties. Both compounds crystallize in the tetragonal *P*4/*nmm* space group. Crystal structure analysis using SPXRD revealed that the Sb and in-plane Se1 sites exhibited large atomic displacement parameters in the in-plane direction. These compounds are n-type semiconductors with optical band gaps of 1.0 eV for $LaOSbSe_2$ and 0.6 eV for $CeOSbSe_2$. The low thermal conductivity of these compounds indicates that they are promising thermoelectric materials upon tuning the carrier density. So far, F substitution has not introduced sufficient electron carriers in these samples. Alternative approaches to increasing the carrier density and electrical conductivity are required to obtain the high-performance thermoelectric properties in this system predicted from first-principles calculations.

**Acknowledgments**

We thank Profs. K. Kuroki and M. Ochi of Osaka University for fruitful discussions. We thank Profs. K. Domen and T. Hisatomi of the University of Tokyo for the preliminary measurements by diffuse reflectance spectroscopy. This work was supported by Grants-in-Aid for Scientific Research (Nos. 15H05886, 16H04493, and 16K17944) and JST-CREST (No. JPMJCR16Q6), Japan.

*E-mail: y_goto@tmu.ac.jp

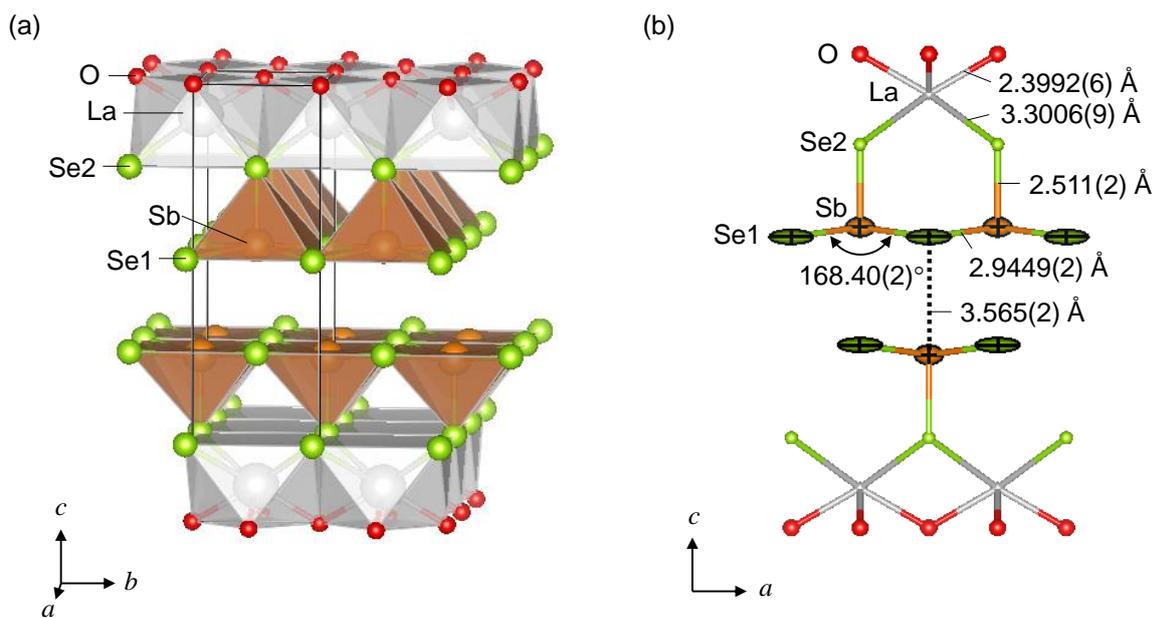

**Fig. 1.** (Color online) (a) Crystallographic structure of LaOSbSe$_2$ (tetragonal *P*4/*nmm* space group). Black lines denote the unit cell. Se ions have two crystallographic sites: the in-plane site (Se1) and out-of-plane site (Se2). (b) Selected bonding distances and angles for LaOSbSe$_2$. For the Sb and Se1 atoms, thermal ellipsoids at the 90% probability level are also shown.

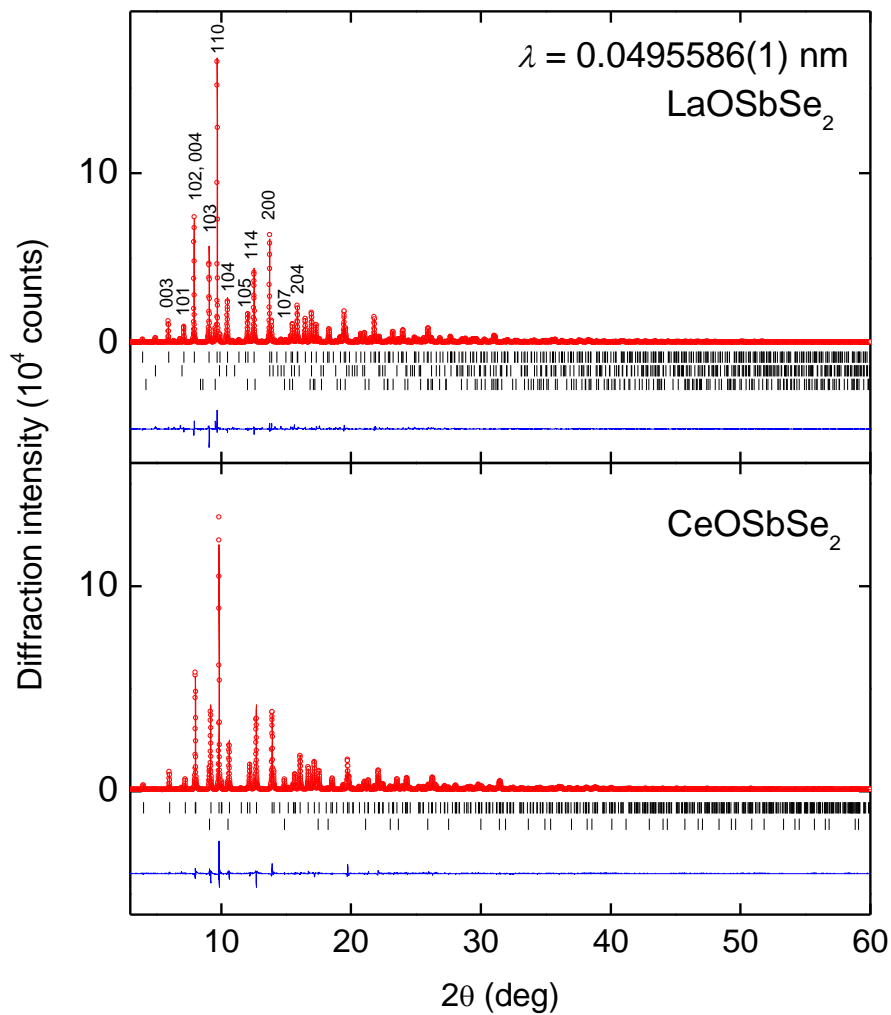

**Fig. 2.** (Color online) Synchrotron powder X-ray diffraction (SPXRD) pattern and results of Rietveld refinement for $LaOSbSe_2$ and $CeOSbSe_2$. The circles and solid curve represent the observed and calculated patterns, respectively, and the difference between the two is shown at the bottom. For $LaOSbSe_2$, the vertical marks indicate the Bragg diffraction positions for $LaOSbSe_2$ (top), Sb (middle), and $La_2O_2Se$ (bottom). For $CeOSbSe_2$, the vertical marks indicate the Bragg diffraction positions for $CeOSbSe_2$ (top) and $CeO_2$ (bottom). Analogous data for F-doped samples are shown in Fig. S1 in the Supporting Information.

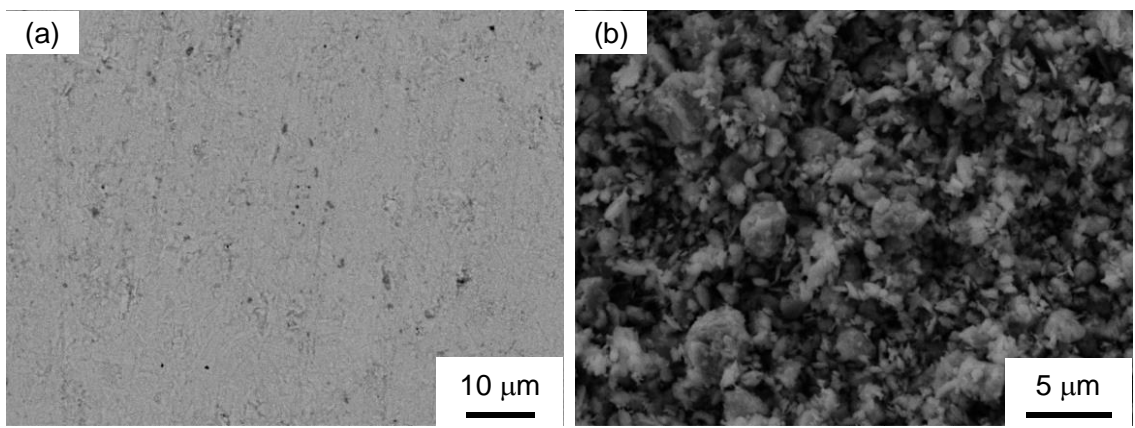

**Fig. 3.** SEM images of (a) polished surface and (b) pulverized powder of $LaOSbSe_2$. Analogous data for other samples are shown in Fig. S2 in the Supporting Information.

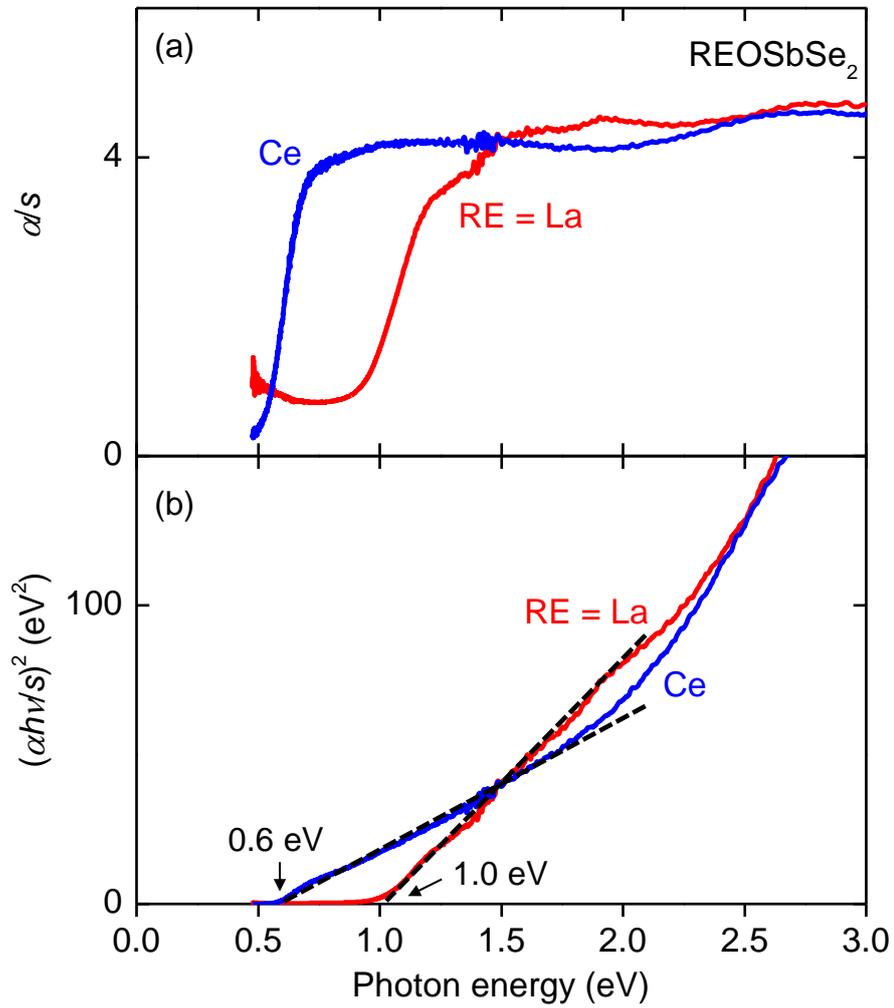

**Fig. 4.** (Color online) (a) $\alpha/s$ and (b) $(\alpha h\nu/s)^2$ plotted as a function of photon energy for LaOSbSe$_2$ and CeOSbSe$_2$. The direct-type absorption edge was evaluated to be 1.0 eV for LaOSbSe$_2$ and 0.6 eV for CeOSbSe$_2$.

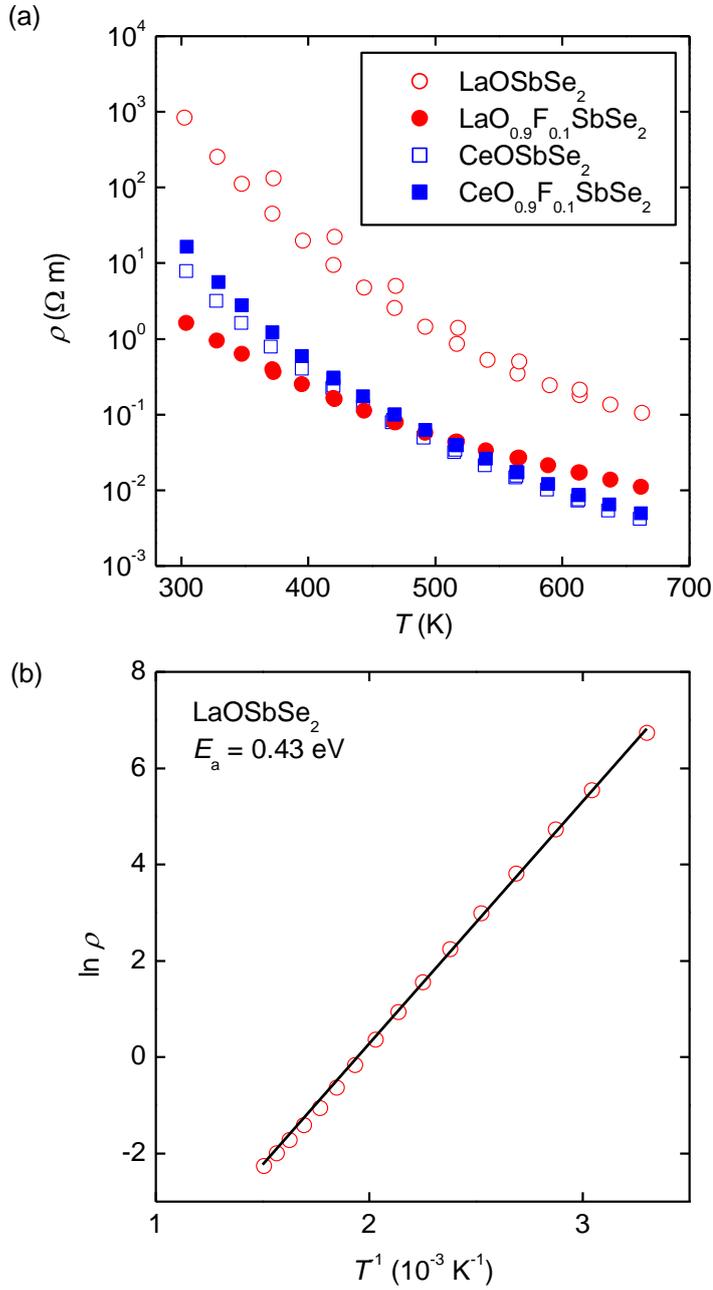

**Fig. 5.** (Color online) (a) Electrical resistivity ($\rho$) as a function of temperature ($T$) for LaO$_{1-x}$F$_x$SbSe$_2$ ($x = 0$ and 0.1) and CeO$_{1-x}$F$_x$SbSe$_2$ ($x = 0$ and 0.1). Measurements in the heating process were performed in 25 K increments, while those in the cooling process were performed in 50 K increments. The measured data in both processes almost coincide except for the case of LaOSbSe$_2$. (b) Arrhenius plot (ln $\rho$ versus $T^{-1}$) of LaOSbSe$_2$. Analogous data for the other samples are shown in Fig. S3.

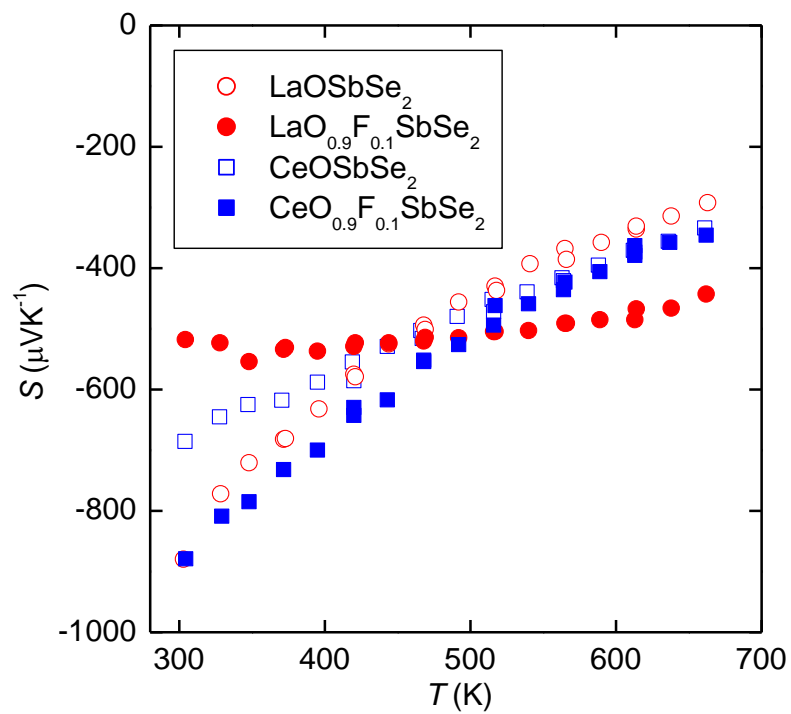

**Fig. 6.** (Color online) (a) Seebeck coefficient ($S$) as a function of temperature ($T$) for $LaO_{1-x}F_xSbSe_2$ ($x = 0$ and $0.1$) and $CeO_{1-x}F_xSbSe_2$ ($x = 0$ and $0.1$).

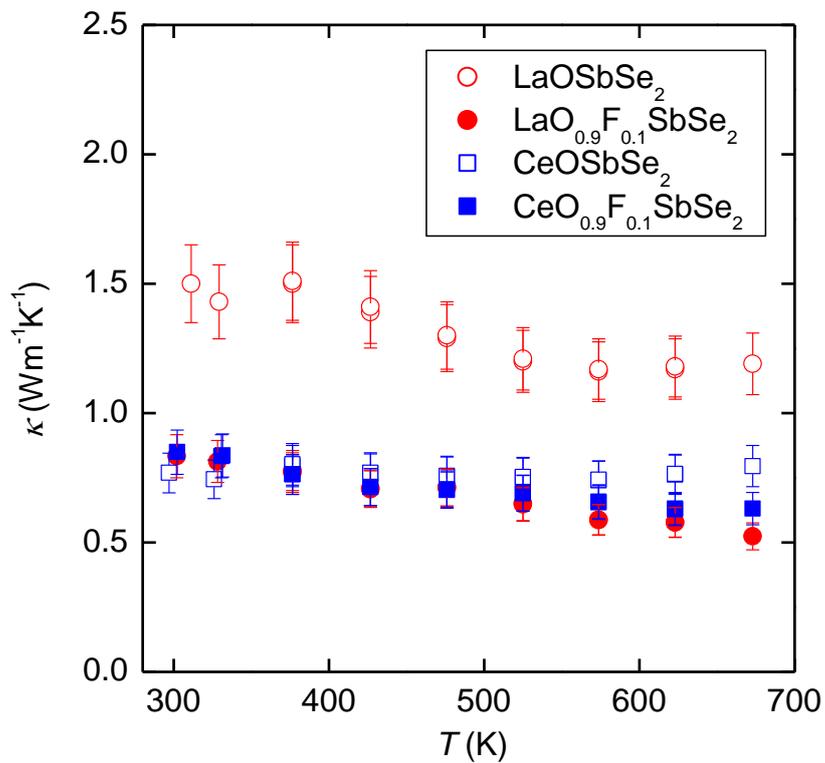

**Figure 7.** (Color online) (a) Thermal conductivity ($\kappa$) as a function of temperature ($T$) for $LaO_{1-x}F_xSbSe_2$ ($x = 0$ and $0.1$) and $CeO_{1-x}F_xSbSe_2$ ($x = 0$ and $0.1$).

**Table I.** Multiplicity and Wyckoff notation (WN), fractional coordinates ($x$, $y$, $z$), and atomic displacement parameters ($U$) refined by Rietveld analysis for LaOSbSe$_2$ and CeOSbSe$_2$.[a,b]

| atom | WN | $x$ | $y$ | $z$ | $U$ ($10^{-4}$ nm$^2$) |
|---|---|---|---|---|---|
| | | | | LaOSbSe$_2$ | |
| La | 2$c$ | 0 | 1/2 | 0.08418(7) | $U_{iso}$ = 0.50(2) |
| O | 2$a$ | 0 | 0 | 0 | $U_{iso}$ = 1.0(2) |
| Sb | 2$c$ | 0 | 1/2 | 0.63741(10) | $U_{11}$ = 2.64(5) |
| | | | | | $U_{33}$ = 1.14(7) |
| | | | | | $U_{eq}$ = 2.1 |
| Se(1) | 2$c$ | 0 | 1/2 | 0.38605(14) | $U_{11}$ = 6.61(13) |
| | | | | | $U_{33}$ = 0.60(13) |
| | | | | | $U_{eq}$ = 4.6 |
| Se(2) | 2$c$ | 0 | 1/2 | 0.80977(13) | $U_{iso}$ = 0.67(4) |
| | | | | CeOSbSe2 | |
| Ce | 2$c$ | 0 | 1/2 | 0.08531(6) | $U_{iso}$ = 0.449(19) |
| O | 2$a$ | 0 | 0 | 0 | $U_{iso}$ = 0.7(2) |
| Sb | 2$c$ | 0 | 1/2 | 0.63179(8) | $U_{11}$ = 2.11(4) |
| | | | | | $U_{33}$ = 1.94(7) |
| | | | | | $U_{eq}$ = 2.0 |
| Se(1) | 2$c$ | 0 | 1/2 | 0.38538(11) | $U_{11}$ = 4.07(10) |
| | | | | | $U_{33}$ = 2.53(13) |
| | | | | | $U_{eq}$ = 3.5 |
| Se(2) | 2$c$ | 0 | 1/2 | 0.81045(10) | $U_{iso}$ = 0.65(3) |

[a]Space group: $P4/nmm$ (No. 129). [b]Values of the last digits in parentheses are standard deviations. For Sb and Se1 sites, anisotropic displacement parameters are refined. Site occupancies are fixed to unity for all sites.

**Table II.** Lattice parameters and reliability factors obtained from Rietveld analysis.[a]

|            | LaOSbSe$_2$   | CeOSbSe$_2$    |
|:----------:|:-------------:|:--------------:|
| $a$ (nm)   | 0.414340(3)   | 0.4808624(3)   |
| $c$ (nm)   | 1.434480(14)  | 1.417697(12)   |
| $R_{wp}$ (%) | 9.908       | 7.655          |
| $R_B$ (%)  | 2.593         | 1.754          |
| GOF        | 3.9362        | 3.0950         |

[a]Values of the last digits in parentheses are standard deviations.

Supporting Information for

# "Synthesis, Crystal Structure, and Thermoelectric Properties of Layered Antimony Selenides REOSbSe$_2$ (RE = La, Ce)"

Yosuke Goto,[1]* Akira Miura,[2] Ryosuke Sakagami,[3] Yoichi Kamihara,[3] Chikako Moriyoshi,[4] Yoshihiro Kuroiwa,[4] and Yoshikazu Mizuguchi[1]

[1]Department of Physics, Tokyo Metropolitan University, 1-1 Minami-osawa, Hachioji, Tokyo 192-0397, Japan

[2] Faculty of Engineering, Hokkaido University, Kita 13, Nishi 8, Sapporo 060-8628, Japan

[3] Department of Applied Physics and Physico-Informatics, Faculty of Science and Technology, Keio University, 3-14-1 Hiyoshi, Yokohama 223-8522, Japan

[4] Department of Physical Science, Hiroshima University, 1-3-1 Kagamiyama, Higashihiroshima, Hiroshima 739-8526, Japan

*E-mail: y_goto@tmu.ac.jp

Table S1. Chemical composition analyzed using energy-dispersive spectroscopy. Standard deviation was calculated using five measurement data.

| Nominal composition | RE : Sb : Se |
|---|---|
| LaOSbSe$_2$ | 0.99(4) : 1 : 1.83(5) |
| LaO$_{0.9}$F$_{0.1}$SbSe$_2$ | 0.97(3) : 1 : 1.90(9) |
| CeOSbSe$_2$ | 1.33(19) : 1 : 2.04(6) |
| CeO$_{0.9}$F$_{0.1}$SbSe$_2$ | 1.14(5) : 1 : 1.8(3) |

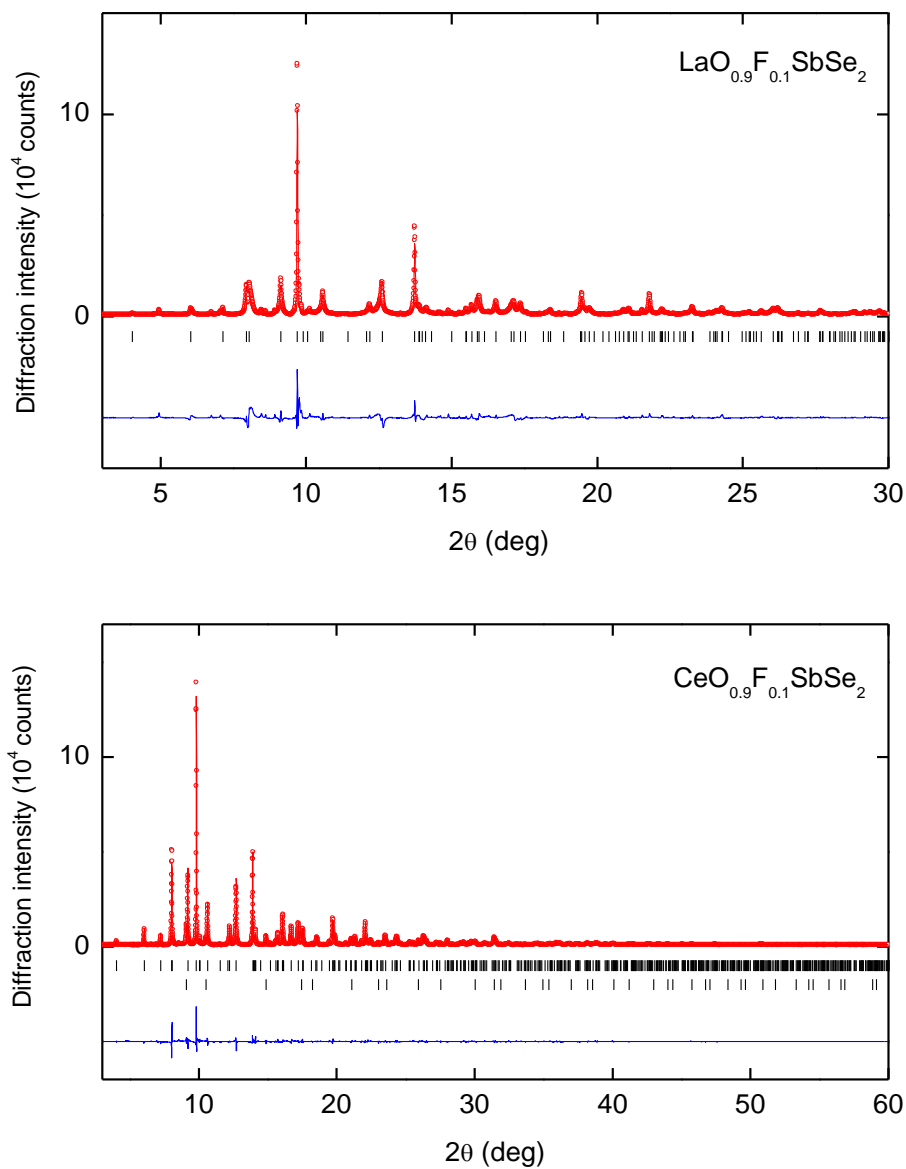

Figure S1. Synchrotron powder X-ray diffraction (SPXRD) pattern and results of Rietveld refinement for LaO$_{0.9}$F$_{0.1}$SbSe$_2$ and Ce O$_{0.9}$F$_{0.1}$SbSe$_2$. The circles and solid curve represent the observed and calculated patterns, respectively, and the difference between the two is shown at the bottom. For LaO$_{0.9}$F$_{0.1}$SbSe$_2$, the vertical marks indicate the Bragg diffraction positions for LaO$_{0.9}$F$_{0.1}$SbSe$_2$. For Ce O$_{0.9}$F$_{0.1}$SbSe$_2$, the vertical marks indicate the Bragg diffraction positions for Ce O$_{0.9}$F$_{0.1}$SbSe$_2$ (top) and CeO$_2$ (bottom). For LaO$_{0.9}$F$_{0.1}$SbSe$_2$, diffraction data up to $2\theta = 30°$ was employed for Rietveld analysis because of broadened diffraction peaks.

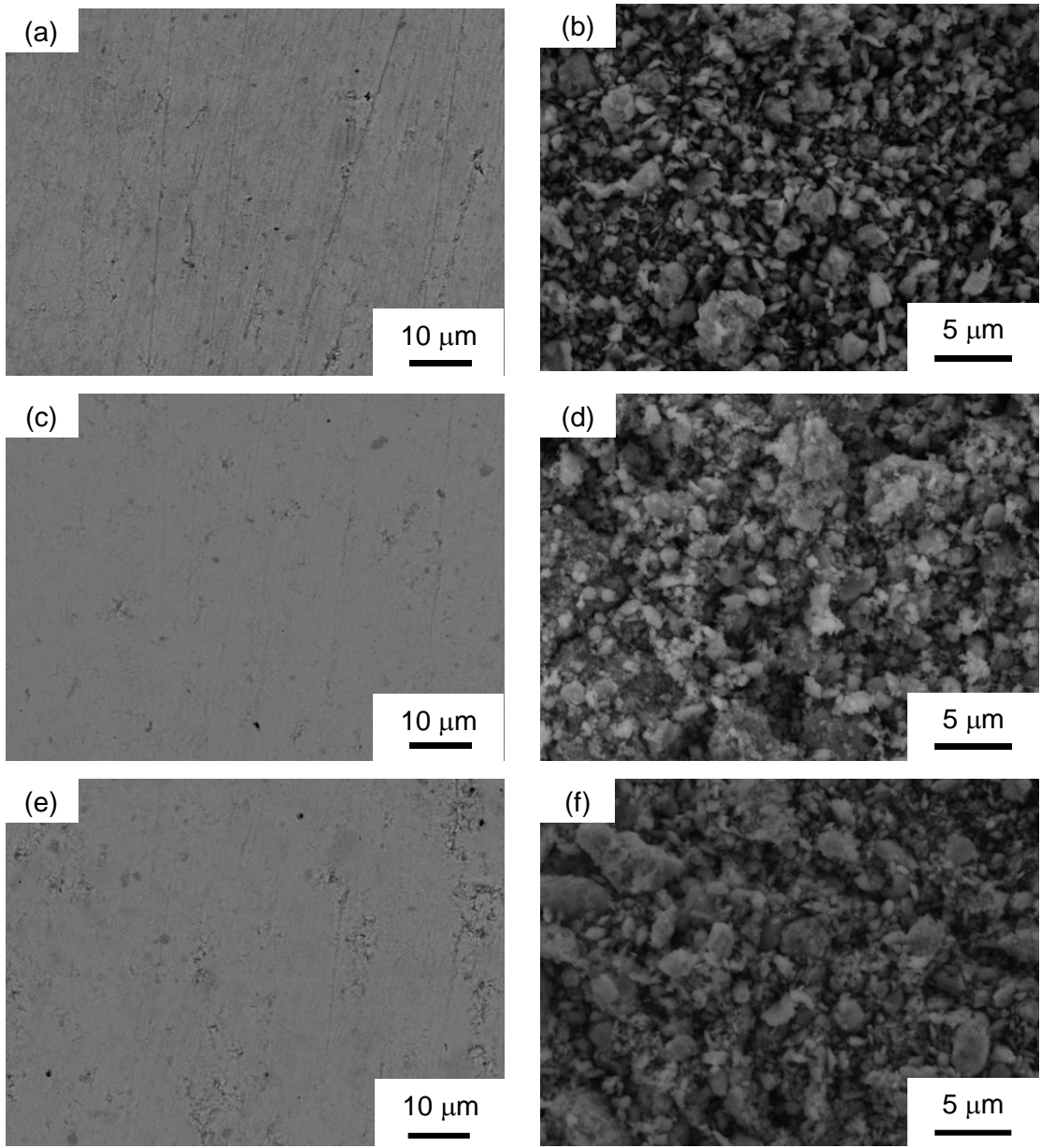

Figure S2. SEM images of (a, c, e) polished surface and (b, e, f) pulverized powder of (a, b) CeOSbSe$_2$, (c, d) LaO$_{0.9}$F$_{0.1}$SbSe$_2$, and (e, f) CeO$_{0.9}$F$_{0.1}$SbSe$_2$.

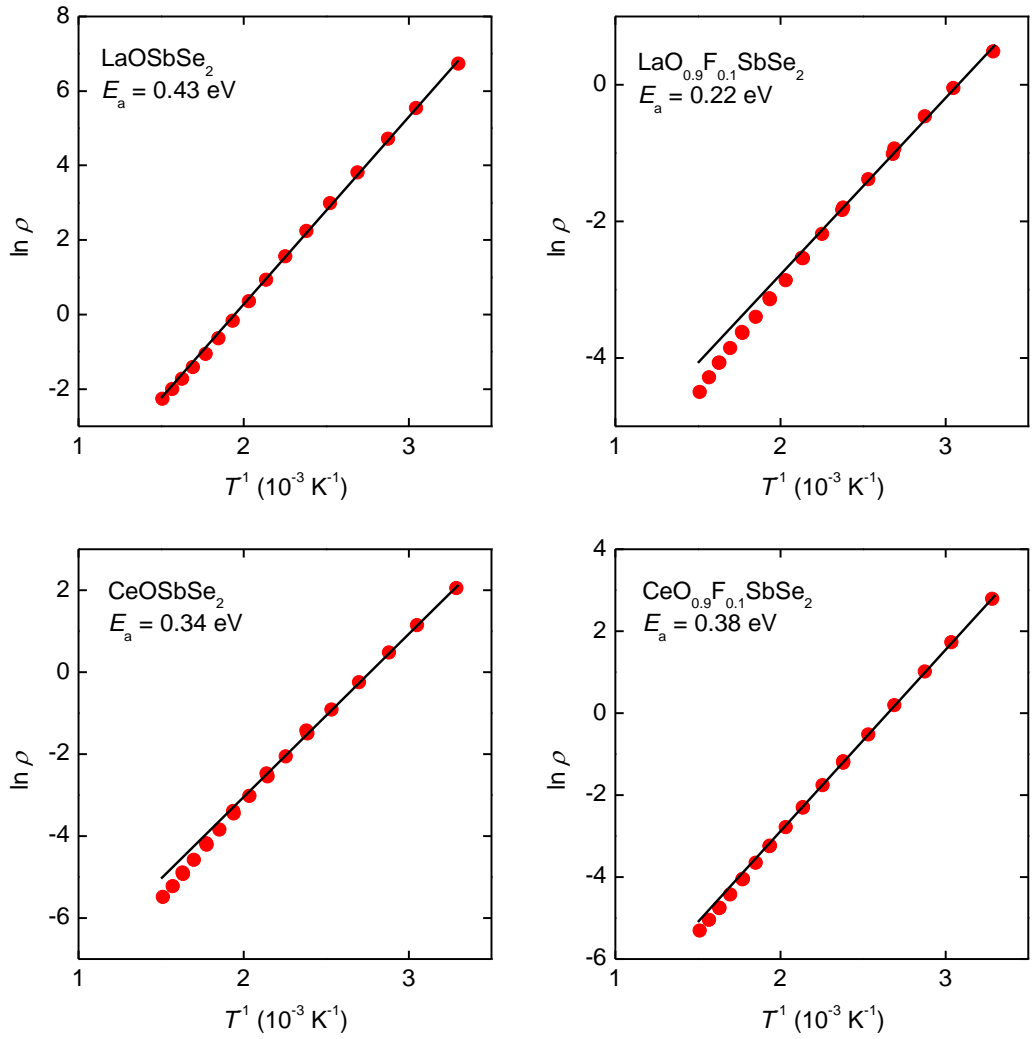

Figure S3. Arrhenius plot, $\ln \rho$ versus $T^{-1}$, for $REO_{1-x}F_xSbSe_2$ (RE = La, Ce; $x = 0, 0.1$).